\documentclass[showpacs,preprint,aps,preprintnumbers,amsmath,amssymb]{revtex4}
\usepackage{amsfonts}
\usepackage{amssymb,amsfonts,graphicx,hyperref,psfrag,indentfirst,epsfig}
\newcommand{\be}{\begin{equation}}
\newcommand{\ee}{\end{equation}}
\newcommand{\bea}{\begin{eqnarray}}
\newcommand{\eea}{\end{eqnarray}}
\newcommand{\ba}{\begin{array}}
\newcommand{\ea}{\end{array}}
\newcommand{\la}{\langle}
\newcommand{\ra}{\rangle}
\let\jnfont=\rm
\def\NPB#1,{{\jnfont Nucl.\ Phys.\ B }{\bf #1},}
\def\PLB#1,{{\jnfont Phys.\ Lett.\ B }{\bf #1},}
\def\EPJC#1,{{\jnfont Eur.\ Phys.\ Jour.\ C }{\bf #1},}
\def\PRD#1,{{\jnfont Phys.\ Rev.\ D }{\bf #1},}
\def\PRL#1,{{\jnfont Phys.\ Rev.\ Lett.\ }{\bf #1},}
\def\MPLA#1,{{\jnfont Mod.\ Phys.\ Lett.\ A }{\bf #1},}
\def\JPG#1,{{\jnfont J.\ Phys.\ G}{\bf #1},}
\def\CTP#1,{{\jnfont Commun.\ Theor.\ Phys.\ }{\bf #1},}

\begin{document}

\title{\ \\[10mm] An Extension for Direct Gauge Mediation of Metastable Supersymmetry Breaking}

\author{\ \\[2mm]  Fuqiang Xu, Jin Min Yang \vspace*{0.5cm} }

\affiliation{Institute of Theoretical Physics, Academia Sinica,
                  Beijing 100190, China  \vspace*{2.5cm} }

\begin{abstract}
We study the direct mediation of metastable supersymmetry breaking
by a $\Phi^2$-deformation to the ISS model and extend it by
splitting both ${\rm Tr}\Phi$ and ${\rm Tr}\Phi^2$ terms in the
superpotential and gauging the flavor symmetry. We find that with
such an extension the enough long-lived metastable vacua can be
obtained and the proper gaugino masses can be generated. Also,
this allows for constructing a kind of models which can avoid the
Landau pole problem. 
{\em 
Especially, in our metastable vacua there exist
a large region for the parameter $m_3$ which can satisfy the 
phenomenology requirements and allow for a low SUSY breaking scale 
($h\mu_2\sim 100$ TeV).
}
\end{abstract}


\maketitle

\section{Introduction}
Dynamical supersymmetry (SUSY) breaking is a convincing scenario to solve
the gauge hierarchy problem, but it seemed an
exceptional phenomenon because its realistic models in general have
to statisfy many theoretical requirements. On the other hand, the
phenomenological considerations are complex when these
dynamical SUSY breaking effects are mediated to the visible sector.

Recently, Intriligator, Seiberg and Shih (ISS)
\cite{Intriligator:2006dd} discovered the meta-stable
supersymmetry breaking in a surprising context of vector-like
theory, which offers a natural framework for dynamical SUSY
breaking and its mediation. Their model (called ISS model) has
received a great deal of attention. However, this model also has
some problems. One is the Landau pole problem which implies that
the unification cannot be simply realized in this model. Another
problem is that the presence of an accidental R-symmetry  (a generic
property of SUSY breaking models) forbids the gaugino
masses. To tackle these probelms, many approaches have been proposed
\cite{Giveon:2007ef,Kitano:2006xg,csaki,Haba:2007rj,dine}.

In the ISS model the mediation of SUSY breaking can proceed through gauge
interaction. Actually, the gauge mediation of dynamical
SUSY breaking was once proposed in \cite{Dine:1981za,Dimopoulos:1981au},
which tries to use a QCD-like strong interaction to break supersymmetry
dynamically and identify the standard model gauge group as a subgroup of
the flavor symmetry. Nevertheless, these early models suffer
from some phenomenological problems, such as the Landau
pole problem and the gaugino mass problem, and thus the idea of gauge
mediation of dynamical SUSY breaking was discarded for a long time.
With the advent of the ISS model, this idea was revived
\cite{Kitano:2006xg,csaki}.

Note that with some deformations the ISS model can give the required phenomenology.
In \cite{Giveon:2007ef} a ${\rm Tr}\Phi^2$ term is introduced to the superpotential
(called $\Phi^2$-deformation) and futher in \cite{Kitano:2006xg} the ${\rm Tr}\Phi$
term in the superpotential is splitted.
In this work we consider a more general deformation to the
ISS model by splitting both ${\rm Tr}\Phi$ and ${\rm Tr}\Phi^2$ terms in the superpotential.
We find that such a general deformation can satisfy the phenomenological constraints
like generating the appropriate gaugino masses and avoiding the Landau pole.

This work is organized as follows.
In Sec. II  we give a brief description for the ISS model and its $\Phi^2$-deformation.
In Sec. III we present our general deformation and discuss its phenomenology.
Finally, in Sec. IV  we give our conclusion.

\section{The ISS model and its $\Phi^2$-deformation}

In the ISS model, the hidden sector is just the ${\cal N}=1$
supersymmetric QCD and has massive quarks with $N_c < N_f
<\frac{3}{2}N_c$, with $N_f$ being the flavor number and $N_c$ the
color number. Its superpotential in the dual magnetic theory is
 \be
 W= hq_i\Phi_{ij}\tilde{q}_j -h\mu^2{\rm Tr}\Phi ,
 \ee
where $q$ and $\tilde{q}$ are respectively the quark and
anti-quark, $\Phi$ is the meson field, and $i$ and $j$ running
from $1$ to $N_f$ are the falvor indices. In the low energy
region, the F-terms of $\Phi$ cannot be simultaneously set to zero
because the rank of $q \tilde{q}$ is $N_f-N_c$ which is smaller
than the rank of $\Phi$ ($=N_f$), and then we get the SUSY-breaking
vaccum energy as
 \be \label{isspot}
 V=N_c|h\mu^2|^2.
 \ee
However, in the high energy range below the scale $\la h\Phi \ra$
the quarks are integrated out and the effective theory is then
$SU(N_f-N_c)$ pure Yang-Mills where the non-perturbative
correction to the superpotential restore the SUSY vacua so that
the SUSY breaking vacua in low energy is only metastable.

The $\Phi^2$-deformation to the ISS model is proposed by Giveon and Kutasov (GK)
\cite{Giveon:2007ef}, whose superpotential
can be derived from brane configuration
\cite{Giveon:2007ew,Ahn:2007si,Ahn:2007vz} and takes the form
 \be \label{m2deformation}
 W= hq_i\Phi_{ij}\tilde{q}_j -h\mu^2{\rm Tr}\Phi
 +\frac{1}{2}h^2\mu_{\phi}{\rm Tr}\Phi^2
 \ee
with $\mu_{\phi}$ being a new energy scale. Note that there is no
non-perturbative superpotential from gaugino condensation effects.
We assume
 \be \label{pert}
 \Phi \leq \frac{\mu^2}{h\mu_{\phi}}
 \ee
to ensure the $\Phi^2$-deformation to be just a perturbation to
the ISS model in the low energy region. This deformed supersymmetric QCD
has a rich landscape of supersymmetric and non-supersymmetric vacua.
The expectation values of fields are given by
  \bea
\langle h\Phi \rangle = \left( \begin{array}{cc}  0 & 0 \\
  0 & \frac{\mu^2}{\mu_{\phi}}I_{N_f-k} \end{array}
 \right) ,
 \quad \langle q\tilde{q} \rangle=\left( \begin{array}{cc} \mu^2I_k & 0\\
 0 & 0
 \end{array} \right),
 \eea
in supersymmetric vacua ($I_n$ denotes a $n\times n$ unit matrix),
and
 \bea \label{metavacua}
\langle h\Phi\rangle= \left( \begin{array}{ccc} 0 & 0 & 0 \\
 0 & h\Phi_n I_n & 0 \\
 0 & 0 & \frac{\mu^2}{\mu_{\phi}}I_{N_f-k-n} \end{array}
 \right) ,
\quad \langle q\tilde{q}\rangle=\left( \begin{array}{ccc} \mu^2I_k & 0 & 0\\
 0 & 0 & 0\\
 0 & 0 & 0 \end{array} \right).
 \eea
in metastable non-supersymmetric vacua.
In the latter case it is necessary to consider the one-loop contribution to
the potential,
which to leading order is given by \cite{Intriligator:2006dd}
 \be
 V_{\rm 1-loop}=b|h^2\mu|^2{\rm Tr}\Phi_n^{\dagger}\Phi_n
 \ee
with $b$ being a constant given by
 \be
 b=\frac{\ln4-1}{8\pi^2}(N_f-N_c).
 \ee
In our following analysis we fix $b=0.01$ for convenience.
The full potential for $\Phi_n,~ q,~ \tilde{q}$ then takes
the form
 \be \label{gkpot}
\frac{V}{|h|^2}=|\Phi_nq|^2+ |\Phi_n\tilde{q}|^2+
|q\tilde{q}-\mu^2I_n +h\mu_{\phi}\Phi_n|^2 +b|h\mu|^2{\rm
Tr}\Phi_n^{\dagger}\Phi_n.
 \ee
To be free of tachyons in dual quark direction and to ensure
the supersymmetric vacua to be far enough from metastable vacua,
we have (neglecting the phase factor in the energy scale and coupling
constants)
 \be \label{tachyonic}
 \frac{\mu_\phi}{\sqrt{b}}\ll \mu \leq \frac{\mu_{\phi}}{bh} .
 \ee
Here the first constraint comes from the requirement of enough-long-lived
metastable vacua (will be discussed in the following), and the second constraint
comes from the special property of this deformation (can also be applied to
our more general deformation discussed in the proceeding section) and can be
obtained from the analysis of the potential in Eq.(\ref{gkpot}) through calculating
$\partial V/\partial \Phi$.
Considering the above constraints, we see that a small $h$ is favored.

To check if the GK metastable vacua is long-lived enough,
we estimate its decay rate by evaluating the Euclidean action
$S_1$ from the GK metastable vacua to their corresponding
true vacua and the action $S_2$ from the GK metastable vacua
to the ISS metastable vacua.
Using the triangle approximation \cite{Coleman:1977py,Duncan:1992ai},
we estimate the bounce action as
 \bea \label{bounce}
&& S_1 \sim \frac{\Delta \Phi^4}{V} \sim \frac{1}{nh^2}
 \left(\frac{\mu_{\phi}/bh}{\mu}\right)^4 ,\\
&& S_2 \sim \frac{\Delta \Phi^4}{V} \sim \frac{1}{nh^6}
\left(\frac{\mu}{\mu_{\phi}}\right)^4,
 \eea
where $\Delta \Phi$ is the corresponding interval between the
two vacua. Therefore, if we have the conditions in Eq.(\ref{tachyonic}),
we can obtain a sufficeintly long lifetime for the universe even if $h$
is not too small.

After building a dynamical supersymmetric breaking model,
the next task is to mediate the breaking effects to the visible
sector. Although the constraints from the long lifetime of the metastable
vacua are weak, the gaugino masses may be much smaller than the sfermion
masses when we use the direct gauge mediation model by
gauging the flavor subgroup in the ISS sector.
In the proceeding section we propose a  more general deformation  by
splitting both ${\rm Tr}\Phi$ and ${\rm Tr}\Phi^2$ terms in the
superpotential. With such a deformation, we can avoid the hierarchy
between gaugino and sfermion masses and, further more, can avoid
the Landau pole problem.

\section{A more general deformation to the ISS model}
In \cite{Kitano:2006xg} the term ${\rm Tr}\Phi$ in the superpotential
is splitted. Here we propose a more general deformation and take the superpotential
as
 \bea \label{xy}
 &W=& h~{\rm Tr}(q\tilde{q}\Phi) -h~\mu_1^2{\rm Tr} Y -h~\mu_2^2{\rm Tr} \hat{\Phi}
 \nonumber \\
&&+\frac{1}{2}h^2 m_1 {\rm Tr} Y^2 +\frac{1}{2}h^2m_2{\rm
Tr}\hat{\Phi}^2
 +h^2m_3{\rm Tr} Z\tilde{Z}
 \eea
with
 \bea
 \Phi= \left( \begin{array}{cc} Y & Z \\
 \tilde{Z} & \hat{\Phi} \end{array} \right) ,
\quad q= \left( \begin{array}{cc} \chi \\
 \rho \end{array} \right).
 \eea
Here $Y$ is a $N_1\times N_1$ matrix, $\hat{\Phi}$ is a $N_2\times N_2$
matrix, and $m_1$, $m_2$, $m_3$, $\mu_1$ and $\mu_2$ are the mass
scales. For $m_1=m_2=m_3=\mu_\phi$ and $\mu_1=\mu_2=\mu$, our deformation
reduces to the GK $\Phi^2$-deformation \cite{Giveon:2007ef}.
The above superpotential has the $SU(N_1)\times SU(N_2)$ flavor symmetry.

Note that our deformation not only exhibits a rich landscape
of supersymmetric and non-supersymmetric vacua just like the
GK $\Phi^2$-deformation, but also has more appropriate phenomenology.
In the following we discuss some features of our deformation.

First, we take a look at the vacua in our deformation.
We get the supersymmetric vacua as
 \bea \label{xyso1}
 \langle h \Phi \rangle =\left( \ba{cccc}0I_k&&& \\ & \frac{\mu_1^2}{m_1}I_{N_1-k} && \\
 && 0I_n & \\ &&& \frac{\mu_2^2}{m_2}I_{N_2-n} \ea \right)  ,
\quad
 \langle q\tilde q\rangle =\left( \ba{cccc} \mu_1^2I_k &&& \\ & 0 && \\&& \mu_2^2I_n & \\
 &&& 0 \ea \right),
 \eea
where $k$ can run from 0 to $N_1$. Considering the one-loop
potential and following the procedure in \cite{Giveon:2007ef}, we
obtain the meta-stable vacua as
 \bea \label{xyso2}
 \langle h\Phi \rangle =\left( \ba{ccc}0I_{N_1}&& \\ & 0I_n & \\ && \frac{m_2}{b}I_{N_2-n} \ea
 \right),
\quad
\langle q\tilde q\rangle= \left( \ba{ccc} \mu_1^2I_{N_1} && \\ & \mu_2^2I_n &\\ && 0 \ea \right) ,
 \eea
where we take $k=N_1$ and the vacuum energy is given to leading
order by
 \be
 V\simeq (N_2-n)|h\mu_2^2|^2.
 \ee
Here the last component of $\Phi$ gives non-zero F-terms.
As discused in the preceding section, we have the condition $m_2/\sqrt{b} \ll \mu_2
\leq m_2/(hb)$ for the long-lived metastable vacua and no tachyonic
quark. Note that in the above we took $k=N_1$. In our following analysis
we also discuss the case of $k=0$ without presenting the explicit structure.
There are, of course, other metastable vacua in our model, but the above vacua
are enough for our purpose of getting an appropriate phenomenological model.

In our meta-stable vacua in Eq. (\ref{xyso2}), the flavor symmetry
$ SU(N_2)$ would be broken to $ SU(n)\times SU(N_2-n)$.
We can gauge the flavor symmetry $SU(n)$ or $SU(N_2-n)$ and
embed the standard model gauge group into the gauged  flavor symmetry
to realize gauge mediation of SUSY breaking.

Now we examine the gaugino masses in our deformation.
The gaugino masses in gauge mediation \cite{dine2,dine3} (with a superpotential
explicitly breaking $R$-symmetry) are given by \cite{Giudice:1997ni}
 \be \label{determine}
 m_{\lambda}= \frac{g^2\bar{N}}{(4
 \pi)^2}F_{X_i}\frac{\partial}{\partial X_i}\log\left({\rm det}{\cal M}\right)
 \ee
where $\bar{N}$ is a constant, ${\cal M}$ is the mass matrix of
messenger fields, and $X_i$ denotes a superfield in the hidden
sector and $-F^*_{X_i}=\partial W/\partial X_i$. In our deformation
the form of ${\cal M}$ is determined by which flavor symmetry,
$SU(n)$ or $SU(N_2-n)$, is gauged:
\begin{itemize}
\item[(1)] If we choose to gauge $SU(N_2-n)$ flavor symmetry and
embed the atandard model group $SU(3)\times SU(2)\times U(1)$ into it,
the messenger fields would be $\rho_2$, $R$, $Z_2$. In our analysis
we use the notation
 \bea
 \Phi=\left( \begin{array}{ccc} Y & Z_1 & Z_2 \\ \tilde Z_1 & \Phi_1 & R\\
 \tilde Z_2 & \tilde R & \Phi_2 \end{array} \right), ~~
 q=\left( \ba{c} \chi \\ \rho_1 \\ \rho_2 \ea \right),
 \eea
where $\Phi_1$ is the $n\times n$ matrix and $\Phi_2$ is the
$(N_2-n)\times(N_2-n)$ matrix. The mass matrix ${\cal M}$ is given by
 \be
 {\cal M}/h=\left( \begin{array}{cccc} \Phi_2 & \mu_1 & 0 & 0 \\
 \mu_1 & hm_3 & 0 & 0 \\ 0 & 0 & \Phi_2 & \mu_2 \\ 0 & 0 & \mu_2 & hm_2 \end{array} \right)
 \ee
in the basis
 \bea
(\rho_2^\prime,~ Z_2,~ \rho_2^{\prime\prime},~ R){\cal M} \left(
 \begin{array}{c} \tilde\rho_2^\prime \\ \tilde Z_2 \\
 \tilde\rho_2^{\prime\prime} \\
 \tilde R \end{array} \right),
 \eea
where $\rho_2$ includes
$\rho_2^\prime$ and $\rho_2^{\prime\prime}$ which couple with
different components of $\chi$.
Therefore, under the assumption $m_2m_3 \gg b \mu_1^2$ we have
 \bea \label{gaginowe}
  m_{\lambda} &=& \frac{g^2\bar{N}}{(4
 \pi)^2}F_{X_i}\frac{\partial}{\partial X_i}\log \left({\rm det}{\cal M}\right) \nonumber \\
 &\simeq & \frac{\alpha}{4 \pi}F_{\Phi_2}\left( \frac{hm_3}{hm_3 \langle \Phi_2 \rangle
  -\mu_1^2} +\frac{hm_2}{hm_2\langle \Phi_2 \rangle-\mu_2^2} \right)\nonumber \\
 &\simeq &\frac{\alpha}{4 \pi} \frac{F_{\Phi_2}}{\langle \Phi_2 \rangle}
 \eea
where $F_{\Phi_2}=h\mu_2^2$ and $\langle \Phi_2 \rangle = m_2/(hb)$ denotes the
expectation value in the meta-stable vacuum shown in Eq.(\ref{xyso2}).
On the other hand, we have the squark masses as
\be
m_s \simeq \frac{\alpha}{4 \pi}
  \left( \frac{F_{\Phi_2}}{\mu_1}+\frac{F_{\Phi_2}}{\langle\Phi_2\rangle} \right)
\simeq \frac{\alpha}{4 \pi}\frac{F_{\Phi_2}}{\langle\Phi_2\rangle}
\ee where we assumed $\mu_1 \gg \langle \Phi_2\rangle$ and
considered $\langle\Phi_2\rangle= m_2/(hb) \gtrsim \mu_2$ as
required by no existence of tachyonic messenger fields. In this
way, we obtain the same order masses for gauginos and squarks.

{\em 
Note that the assumptions $m_2m_3 \gg b \mu_1^2$ and $\mu_1 \gg \langle
\Phi_2\rangle$ are easy to be satisfied if we let $m_3$ large
enough. We checked that these conditions do not affect our vacua
structure in Eqs.(\ref{xyso1},\ref{xyso2}) and also do not affect 
our calculations about the lifetime of the vacua (in our calculations 
we use Eq. \ref{bounce} and replace $\mu_\phi$ and $\mu$ with $m_2$ 
and $\mu_2$, respectively). 
Compared with \cite{Kitano:2006xg}, where the two independent scales 
are required to be nearly equal, i.e., $m_3\sim \mu_1$, in our study 
we have a larger appropriate region for the parameter $m_3$.
Actually, as shown from our following analysis of Landau pole, 
a large $m_3$ is favored, which enables us to obtain a SUSY
breaking scale ($h\mu_2 \sim 100$ TeV) lower than the value 
obtained in \cite{Kitano:2006xg}.
}

\item[(2)] If we choose to gauge the $SU(n)$ flavor symmetry and embed
the standard model group into it, the messenger fields would be
$\rho_2$, $\chi$, $R$ and $Z_1$. Then the mass matrix ${\cal M}$ is given by
 \be
 {\cal M}/h=\left( \begin{array}{cccc} \Phi_2 & \mu_2 & 0 & 0 \\
 \mu_2 & hm_2 & 0 & 0 \\ 0 & 0 & Y & \mu_2 \\ 0 & 0 & \mu_2 & hm_3 \end{array} \right)
 \ee
in the basis
 \bea
 (\rho_2,~ R,~ \chi,~ Z_1){\cal M} \left(
 \begin{array}{c} \tilde\rho_2 \\ \tilde R \\
 \tilde\chi \\
 \tilde Z_1 \end{array} \right).
 \eea
Therefore, if we assume the F-term of $Y$ field is not
zero (for example, we can take $k=0$ and a non-zero $N_1$, and then in
Eq. \ref{xyso2} the first diagonal element is $(m_1/b) I_{N_1}$ for $\langle h\Phi\rangle$
and $0 I_{N_1}$ for $\langle q\tilde q\rangle$ ) and further assume $m_1m_3 \gg b \mu_2^2$,
we have
 \bea \label{gaginowe}
  m_{\lambda} &=& \frac{g^2\bar{N}}{(4
 \pi)^2}F_{X_i}\frac{\partial}{\partial X_i} \log \left({\rm det}{\cal M}\right)  \nonumber \\
 &\simeq & \frac{\alpha}{4 \pi}\left( h^2m_2 +\frac{F_{Y} hm_3}{hm_3\langle Y\rangle-\mu_2^2}
 \right)\nonumber \\
 &\simeq &\frac{\alpha}{4 \pi} \frac{F_{Y}}{\langle Y\rangle}
 \eea
where $F_Y=h\mu_1^2$ and  $\langle Y  \rangle = m_1/(hb)$ denotes
the expectation value in the meta-stable vacuum. On the other hand,
we have the squark masses as \be m_s\simeq \frac{\alpha}{4
\pi}\left(\frac{F_{\Phi_2}}{\langle \Phi_2 \rangle}
  +\frac{F_{Y}}{\langle Y\rangle}\right)
\ee
Therefore, in this case if $\mu_1\sim \mu_2$
the squark masses can also be of the same order as gaugino
masses.
\end{itemize}

Finally, we check if our deformation is free of the Landau pole problem.
The mass spectrum can be read out from the metastable vacua and is dependent on which
flavor symmetry, $SU(n)$ or $SU(N_2-n)$, is gauged.
We found that our model has the Landau pole problem
if we choose to gauge the $SU(n)$ symmetry, but is free of the Landau pole
problem if we choose to gauge the $SU(N_2-n)$ symmetry. In the following
we demonstrate how to avoid the Landau pole problem in case of
gauging the $SU(N_2-n)$ symmetry.

{\em 
When we gauge $SU(N_2-n)$ and embed the standard model group into
it, the $\rho_2^\prime$ and $Z_2$ have a mass of ${\cal O}(m_2/b)$ and
${\cal O}(hm_3)$, respectively. }
The $R$ and $\rho_2^{\prime\prime}$ have a
mass near the scale $h\mu_2$, and the pseudo-moduli $\Phi_2$ has
a mass of similar size to the gauginos. In our following calcualtion
we take $m_2/b\sim h\mu_2$ for simplicity. The beta function
coefficients of the $SU(3)$ gauge coupling $b_3$ is given by
 \bea \label{b3}
 &&b_3(\mu_R<m_\lambda)=b_3^{SM}=-7, \nonumber \\
 &&b_3(m_\lambda < \mu_R < h\mu_2)=-3+N_2-n, \nonumber \\
 &&b_3(h\mu_2< \mu_R < hm_3)=-3+N_2+N_f-N_c, \nonumber \\
 &&b_3(hm_3< \mu_R < \Lambda)=-3+2N_f-N_c, \nonumber \\
 &&b_3(\mu_R>\Lambda)=-3+N_c,
 \eea
where $\mu_R$ is the renormalization scale. In our analysis we use
the definition
 \be
 \mu_R\frac{dg_3}{d\mu_R}=\frac{b_3}{16\pi^2}g_3^3\equiv\beta_3,
 \ee
and take the input parameters as
 \be
 M_{GUT}= 10^{16}~{\rm GeV},~M_z \simeq 90~{\rm GeV},~m_\lambda \simeq
 10^3~{\rm GeV}, ~\frac{g_3^2(M_z)}{4\pi} \sim 0.18.
 \ee
The $SU(3)$ coupling is obtained as
 \bea
 \label{alphas}
 \alpha^{-1}_3(M_{GUT}) &\simeq&  5.6 - \frac{7}{2\pi}\log M_z +\frac{4+N_2-n}{2\pi}\log
      m_\lambda + \frac{N_f-N_c+n}{2\pi}\log (h\mu_2) \nonumber  \\
 && +\frac{N_f-N_2}{2\pi}\log (hm_3) +\frac{2N_c-2N_f}{2\pi}\log
 M_{GUT},
 \eea
where we take $\Lambda=M_{GUT}$. 
{\em 
For example, taking $N_2-n=5$,
$N_f=11$, $N_c=9$, $n=1$ and $N_1=N_f-N_2=5$, 
we find that for the SUSY
breaking scale $h\mu_2 \sim 10^5$ GeV and $hm_3\geq 10^7$ GeV,
the Landau pole can be avoided below the unification scale, i.e.
$\alpha^{-1}_3(M_{GUT})>1$. Under this condition, proper gaugino
mass can be obtained from Eq.(\ref{gaginowe}), and we checked it 
is consistent with our other assumptions.
}

\section{Conclusion}
In this work we considered a more general
deformation to the ISS model by splitted both ${\rm Tr}\Phi$ and ${\rm Tr}\Phi^2$.
Then we found that the corresponding metastable
vacua can be enough long-lived and the proper gaugino masses can
be generated. In particular, there can exist a kind of models
which can avoid the Landau pole problem if we gauge $SU(N_2-n)$
flavor group and embed the standard model group into it.

\section*{Acknowledgement}
We thank R. Kitano, Z.F. Kang, G. Yang and W.S. Xu for useful
discussions. This work was supported in part by National Natural
Science Foundation of China (NNSFC) under number Nos.  10725526 and 10635030.

\end{document}